\documentclass[prb,twocolumn,superscriptaddress,showpacs,epsf]{revtex4}

\usepackage{graphicx}
\usepackage{latexsym}
\usepackage{amssymb}
\usepackage{amsfonts}
\usepackage{soul}
\usepackage{color}
\begin{document}
\bibliographystyle{apsrev}

\title{Reverse--domain superconductivity in superconductor--ferromagnet hybrids: \\
effect of a vortex--free channel on the symmetry of $I-V$
characteristics }

\author{A.Yu. Aladyshkin}
\email{aladyshkin@ipm.sci-nnov.ru} \affiliation{INPAC -- Institute
for Nanoscale Physics and Chemistry, Nanoscale Superconductivity
and Magnetism
\\ and Pulsed Fields Group, K.U.~Leuven, Celestijnenlaan 200D,
B--3001 Leuven, Belgium} \affiliation{Institute for Physics of
Microstructures, Russian Academy of Sciences, 603950 Nizhny
Novgorod, GSP-105, Russia}
\author{D.Yu. Vodolazov}
\affiliation{Institute for Physics of Microstructures, Russian
Academy of Sciences, 603950 Nizhny Novgorod, GSP-105, Russia}
\author{J. Fritzsche}
\affiliation{INPAC -- Institute for Nanoscale Physics and
Chemistry, Nanoscale Superconductivity and Magnetism \\ and Pulsed
Fields Group, K.U.~Leuven, Celestijnenlaan 200D, B--3001 Leuven,
Belgium}
\author{R.B.G. Kramer}
\affiliation{INPAC -- Institute for Nanoscale Physics and
Chemistry, Nanoscale Superconductivity and Magnetism \\ and Pulsed
Fields Group, K.U.~Leuven, Celestijnenlaan 200D, B--3001 Leuven,
Belgium}\affiliation{Institut N\'eel, CNRS--Universit\'e Joseph Fourier, BP 166, 38042 Grenoble Cedex 9, France}
\author{V.V. Moshchalkov}
\affiliation{INPAC -- Institute for Nanoscale Physics and
Chemistry, Nanoscale Superconductivity and Magnetism \\ and Pulsed
Fields Group, K.U.~Leuven, Celestijnenlaan 200D, B--3001 Leuven,
Belgium}

\date{\today}
\begin{abstract}
We demonstrate experimentally that the presence of a single domain
wall in an underlying ferromagnetic BaFe$_{12}$O$_{19}$ substrate
can induce a considerable asymmetry in the current ($I$) --
voltage ($V$) characteristics of a superconducting Al bridge. The
observed diode--like effect, i.e. polarity--dependent critical
current, is associated with the formation of a vortex--free
channel inside the superconducting area which increases the total
current flowing through the superconducting bridge without
dissipation. The vortex--free region appears only for a certain
sign of the injected current and for a limited range of the
external magnetic field.
\end{abstract}



\maketitle

The development of material deposition techniques and lithographic
methods have made it possible to fabricate
superconductor--ferromagnet (S/F) hybrid structures with
controlled arrangements of ferromagnetic layers/elements.
\cite{Martin-JMMM-03,Velez-JMMM-08,Aladyshkin-SuST-09} These
flux-- and exchange--coupled S/F hybrids
\cite{Velez-JMMM-08,Aladyshkin-SuST-09,Lyuksyutov-AdvPhys-05,Buzdin-RMP-05}
are of fundamental interest for investigations of nontrivial
interactions between superconductivity and nonuniform
distributions of magnetization. In addition S/F hybrids seem to be
potential candidates for the development of tunable elements of
superconducting electronics.\cite{Aladyshkin-SuST-09}

It is known that a nonuniform magnetic field can modify the
conditions for the appearance of superconductivity due to the
effect of a local field
compensation.\cite{Aladyshkin-PRB-03,Lange-PRL-03} In flux-coupled
S/F bilayers,\cite{Aladyshkin-SuST-09} the formation of localized
superconductivity results in either domain--wall superconductivity
(DWS) or reverse--domain superconductivity (RDS), when
superconductivity occurs, respectively, above magnetic domain
walls or above magnetic domains of opposite polarity with respect
to the orientation of an external magnetic field $H_{ext}$ (see
review \cite{Aladyshkin-SuST-09} and references therein). The
appearance of localized superconductivity (DWS and RDS) becomes
possible if the amplitude of the nonuniform field, $B_0$, is
comparable or exceeds the upper critical field, $H_{c2}$, of the
superconducting material, which was confirmed experimentally for
various planar S/F structures.
\cite{Gillijns-PRL-05,Yang-NatMat-04,Yang-PRB-06,Yang-ARL-06,Fritzsche-PRL-06,Aladyshkin-APL-09}

A present challenge, associated with these S/F hybrids, is the
direct investigation of the transport properties of
superconducting channels that are induced by stray magnetic
fields. Indeed, this problem seems to be crucial for any practical
applications exploring the effect of localized superconductivity
and guided vortex motion in tunable magnetic landscapes. Parallel
magnetic domains in thick permalloy films were found to lead to a
preferential vortex motion and a giant anisotropy of the critical
currents in S films and crystals,
\cite{VlaskoVlasov-PRB-08a,Belkin-APL-08,Belkin-PRB-08,VlaskoVlasov-PRB-08b,Belkin-APL-10}
even though the amplitude of the nonuniform field appears to be
insufficient for the formation of localized superconductivity
($B_0/H_{c2}<1$ at low temperatures).

It is important to note that the magnetic field induced by
parallel magnetic domains in BaFe$_{12}$O$_{19}$ is rather high
and thus suitable for RDS in superconducting Al films since
$B_0/H_{c2}>2.5$ for all temperatures.\cite{Aladyshkin-APL-09}
Using such S/F bilayers with well defined localized
superconducting channels in a normal-metal matrix, we continue our
previous study \cite{Aladyshkin-APL-09} with the aim to test the
potential of the superconducting channels to carry current. In
this Letter we focus on the measurements of the current ($I$) --
voltage ($V$) characteristics of a S/F bilayer along a {\it
single} domain wall in a ferromagnetic substrate as a function of
$H_{ext}$ and analyze the dependence of the critical current $I_c$
on $H_{ext}$ for different signs of the bias current.

    \begin{figure}[tb]
    \begin{center}
    \includegraphics[width=7.5cm]{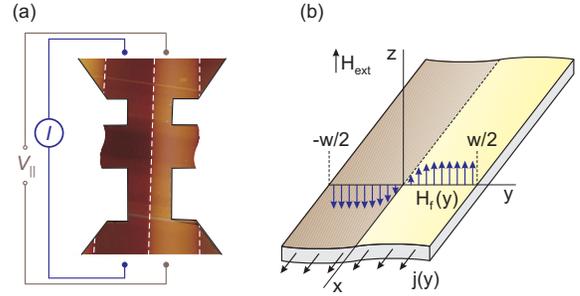}
    \end{center}
    \caption{(color online) (a) Combined magnetic and atomic force microscopy
    images of the hybrid sample. The dark and bright stripes, corresponding to the
    different ferromagnetic domains, are shown only within
    the cross-shaped superconducting bridge. White dashed lines depict the position of the domain walls.
    All elements of the electrical circuit is shown schematically.
    (b) The model S/F structure.}
    \label{Fig-System2}
    \end{figure}

Our sample consists of a ferromagnetic crystal BaFe$_{12}$O$_{19}$
with a thin-film superconducting Al bridge grown on top. Since the
ferromagnetic and superconducting parts were electrically isolated
by a 5 nm Si buffer layer, the interaction between these parts was
purely magnetostatic. Being cut along the proper crystallographic
direction, the polished crystal BaFe$_{12}$O$_{19}$ exhibits a
stripe-type domain structure with dominant in-plane
magnetization.\cite{Aladyshkin-APL-09} The location of the domain
walls was determined by magnetic force microscopy, prior to the
preparation of the superconducting bridge. The cross-shaped Al
microbridge (30 $\mu$m wide and 50 nm thick) was fabricated by
e-beam lithography, molecular beam epitaxy and lift-off etching
[Fig. \ref{Fig-System2}(a)]. A similar structure was used in Ref.
\cite{Aladyshkin-APL-09} for the observation of the anisotropy of
the electrical resistance in this S/F bilayer.

    \begin{figure}[tb]
    \begin{center}
    \includegraphics[width=8.0cm]{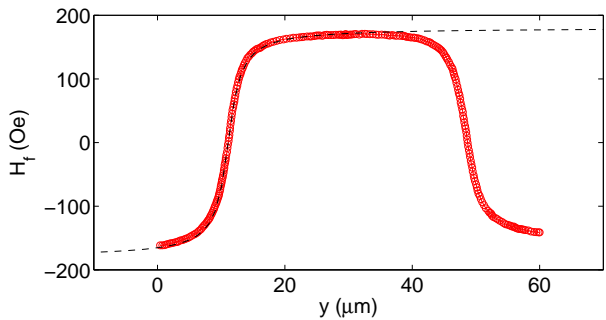}
    \end{center}
    \caption{(color online) The $z-$component of the field induced
    by the ferromagnetic domains, measured by a scanning Hall
    probe microscope along a line perpendicular to the domain walls at $T=72$~K, $H_{ext}=0$ and at height 400~nm.
    The dashed line corresponds to $H_f=H_0\,\arctan\,(y/L)$ (\mbox{$H_0=115$~Oe}, $L=1.5~\mu$m).}
    \label{Fig-Field2}
    \end{figure}

The profile of the perpendicular $z-$component of the nonuniform
magnetic field is shown in Fig. \ref{Fig-Field2}. Its amplitude,
even though measured at a rather large distance (400 nm) from the
surface, is close to the $H_{c2}$ value for Al films
($H_{c2}\simeq 2\cdot10^2$~Oe at $T=0$). Since the field amplitude
inside the superconducting film, $B_0$, will exceed $H_{c2}$ at
all temperatures, superconductivity appears in this S/F system
only at $H_{ext}\neq 0$ as reverse-domain superconductivity above
positively (negatively) magnetized domains at $H_{ext}<0$
($H_{ext}>0$), respectively.

\begin{figure}[tb]
    \begin{center}
    \includegraphics[width=8.5cm]{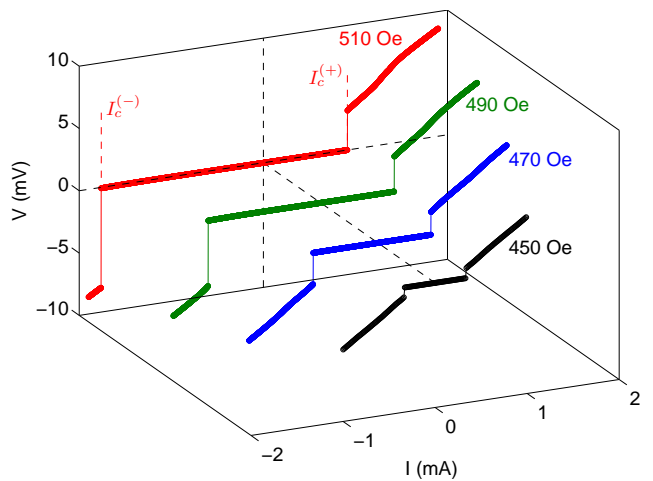}
    \end{center}
    \caption{(color online) Typical $I-V$ dependencies measured along the
    domain wall for different $H_{ext}$ values. Both branches for $I>0$ and $I<0$ were measured starting from
    $I=0$. \label{Fig-IV-examples}}
    \end{figure}

Figure~\ref{Fig-IV-examples} shows the typical $I-V$
characteristics measured at $T=0.5$~K ($T/T_{c0}\simeq 0.34$).
Depending on the $H_{ext}$ value, there are three different cases:

(i) symmetric normal--type $I-V$ dependence with almost constant
slope $dV/dI$ (not shown here);

(ii) symmetric $I-V$ dependence with non-zero critical current
(curves labelled 450 Oe and 470 Oe). This is realized in a rather
wide $H_{ext}$ range corresponding to the reverse-domain
superconductivity;

(iii) asymmetric hysteretic $I-V$ dependence (curves labelled 490
Oe and 510 Oe) with $I^{(+)}_{c}\neq I^{(-)}_{c}$. Here we
introduce the critical currents $I^{(+)}_{c}$ and $I^{(-)}_{c}$
for the ascending branches of the $I-V$ curves both for positive
($+$) and negative ($-$) polarities of the transport current. This
type of $I-V$ characteristics was found only in the close vicinity
of the compensation field ($|H_{ext}|\simeq B_0$).

\begin{figure}[tb]
    \begin{center}
    \includegraphics[width=8.5cm]{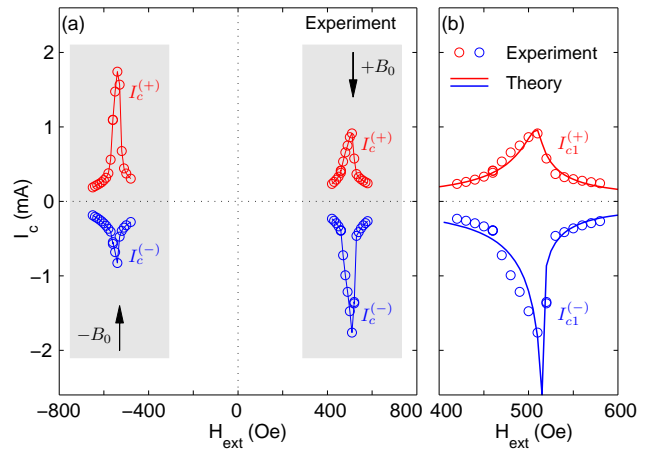}
    \end{center}
    \caption{(color online) (a) Critical currents $I^{(\pm)}_{c}$  as a function of $H_{ext}$ at $T=0.5$~K.
    The shaded areas correspond to the $H_{ext}$ range
    where the decrease of the resistance from its normal value (i.e. the RDS regime) was detected.
    Arrows indicate the compensation field $|H_{ext}|=B_0$.
    (b) Comparison between experiment (circles) and theory (solid lines).} \label{Fig-CriticalCurrents}
    \end{figure}

The dependencies of the critical currents $I^{(\pm)}_{c}$ on
$H_{ext}$ are summarized in Fig.~\ref{Fig-CriticalCurrents}(a).
The relationship between $I^{(+)}_{c}$ and $I^{(-)}_{c}$ depends
both on the absolute value of $H_{ext}$ and its sign:
$I^{(+)}_{c}>I^{(-)}_{c}$ at $H_{ext}<0$ and vice versa. Thus, the
most important finding of this paper is the field-induced change
of the symmetry of the $I-V$ characteristics. The fact that the
transmission capacity of the superconducting channel formed in the
non-uniform magnetic field, can be strongly dependent on the
polarity of the transport current may look rather unusual and
counter-intuitive.

    \begin{figure}[tb]
    \begin{center}
    \includegraphics[width=8.8cm]{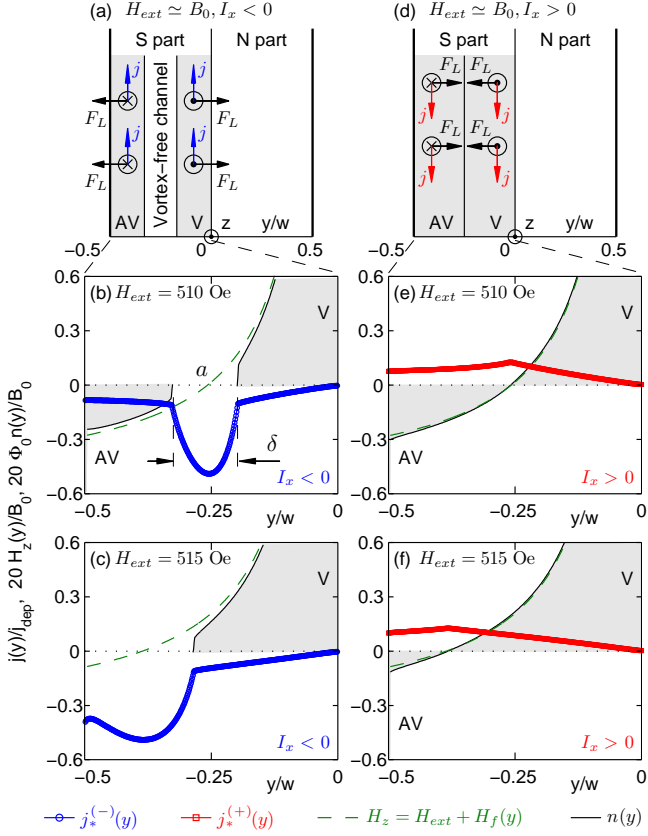}
    \end{center}
    \caption{(color online) Static vortex and current patterns in the S bridge
    in the presence of the non-uniform field
    $H_f=H_0\,\arctan\,(y/L)$, corresponding to the maximum of the
    flowing current, at different $H_{ext}$ values. V and AV stand for vortices and
    antivortices, $j_{dep}$ is the depairing current density \cite{Comment}.
    The right part of the superconducting bridge ($0<y<w/2$) is in the normal state at $H_{ext}>0$ and $j=0$
    and $n=0$ there.} \label{Fig-Pattern}
    \end{figure}

To clarify the physical origin of the difference between
$I_{c}^{(+)}$ and $I_{c}^{(-)}$, we analyze the transport
properties of the generic S/F bilayer within the London model. We
consider a thin superconducting bridge (width $w$) in the field of
a single straight domain wall positioned in the center of the
bridge \mbox{($y=0$)} and aligned parallel to the $x-$axis
[Fig.~\ref{Fig-System2}(b)]. We assume that all parameters of the
resulting current/vortex structures depend only on the transverse
$y-$coordinate. The current density $j(y)$ and vortex density
$n(y)$ in the presence of a nonuniform field meet the
Maxwell--London equation
\cite{Vodolazov-PhysC-2001,Vodolazov-PRB-2005}
    \begin{eqnarray}
    \nonumber \frac{4\pi\lambda^2}{c} \frac{dj}{dy} +
    \frac{d}{c}\int\limits_{-w/2}^{w/2}
    \frac{j(y^{\prime})\,dy^{\prime}}{y^{\prime}-y} =
    H_{ext} + H_{f}(y) - \Phi_0 n(y),
    \label{Eq-MaxwellLondon}
    \end{eqnarray}
where $H_f(y)=H_0\,{\rm arctan}(y/L)$ is the field induced by the
domain wall, $\lambda$ is the London penetration depth, $d$ is the
thickness of the superconductor, $\Phi_0=\pi\hbar c/e$ is the flux
quantum. To determine the critical current $I_c$ of the
superconducting strip, corresponding to the transition between the
state with motionless vortices to the flux motion regime, we apply
the following conditions:

(i) the maximum of the current density in the vortex-free region
should be equal to the critical current density $j_s$, which
defines the threshold value for the nucleation of vortices and
antivortices inside the superconductor or at its edges;

(ii) in order to guarantee a flux motion regime at \mbox{$I=I_c$},
$j(y)$ should be equal to the depinning current density $j_p$ in
the area where $n(y)\neq 0$, with $j_p=j_{p0}/(1+|B_z|/B_p)$
    \begin{eqnarray}
    \nonumber j_p=\frac{j_{p0}}{(1+|B_z|/B_p)}
    \end{eqnarray}
(according to Kim-Anderson model\cite{Campbell_Evetts}) and $B_z$
    \begin{eqnarray}
    \nonumber B_z = H_{ext} + H_{f} + \frac{d}{c}\int\limits_{-w/2}^{w/2}
    \frac{j(y^{\prime})\,dy^{\prime}}{y-y^{\prime}}
    \end{eqnarray}
is the local magnetic field. As a result, in the vortex--free
region, $n(y)=0$, the current density can be larger than $j_p$.

(iii) the profile $j_*(y)$, which satisfies both conditions (i)
and (ii), allows us to define the critical current as follows
$I_c=\int_{-w/2}^{w/2} j_*(y)\,dy$.

In our calculations we use the parameters typical for our system:
$w=30~\mu$m, $d=50~$nm,$\lambda=150~$nm, $B_0=520~$Oe,
$B_p=30~$Oe, $H_0=331~$Oe, $j_{p0}=0.14\,j_{dep}$,
$j_s=0.55\,j_{dep}$, where $j_{dep}$ is the depairing current
density of Al at low temperatures.\cite{Romijn} Our choice for the
parameter $L=0.35~\mu$m seems to be reasonable since the width of
the transient area in the $B_z$--distribution inside the
superconducting film (at $h<50~$nm) can be substantially smaller
than that measured at large distances ($L=1.5~\mu$m at
$h=400~$nm). Since there is rather good agreement between the
experimental data and the calculated dependencies
$I_c^{(+)}(H_{ext})$ [Fig.~\ref{Fig-CriticalCurrents}(b)], we can
interpret the asymmetry of the transmission capacity of the
superconducting channels as follows.

Provided $H_{ext}\simeq B_0$, the magnetic field is effectively
compensated in the left part of the bridge, while the right part
will be switched to the normal state. Since the local field
$H_z=H_{ext}+H_f$ changes its sign inside the superconducting area
(in the case $H_{ext}<B_0$), the stable vortex structure should
generally consist of vortices and antivortices. However the
Lorentz force \mbox{${\bf F}_L=c^{-1}\,[{\bf j}\times {\bf
\Phi}_0]$} acting on a vortex depends on the direction of the
transport current $j$, therefore the resulting vortex pattern,
corresponding to the non-dissipative current flow, may be
dependent on the sign of $I_x$. Indeed, an injection of the
negative bias current ($I_x<0$) forces vortices (antivortices) to
move to the right (left), resulting in a vortex--free channel
inside the RDS area [Fig.~\ref{Fig-Pattern} (a1)--(c1)]. The exact
position $a$ and the width $\delta$ of such channel depends on
$H_{ext}$. Since there are no vortices in a certain area, one can
apply a larger current through such a vortex-free channel without
loosing energy and the excess current due to this effect can be
roughly estimated as \mbox{$\delta\times(j_s-j_{p0})$}. However,
the vortex-free area is absent for a bias current of opposite
polarity ($I_x>0$), since vortices and antivortices move in
counter directions and annihilate at the point of zero magnetic
field [Fig.~\ref{Fig-Pattern} (a2)--(c2)]. In this case the
current density cannot be larger than $j_{p0}$ and
$I_c^{(+)}<I_c^{(-)}$. In the under-compensated regime, when the
absolute $H_{ext}$ value is substantially less than $B_0$, the
gradient of the local field $\left(dH_z/dy\right)_a$ increases
rapidly as $H_{ext}$ decreases [compare Fig.~\ref{Fig-Pattern}
(a1)--(c1)], $\delta\to 0$, and the diode effect vanishes. In the
over-compensated regime ($H_{ext}>B_0$) the vortex--free region
positioned near the left edge of the bridge becomes very narrow
($\delta\to 0$) and, as a consequence, the excess current goes to
zero and the symmetry of the $I-V$ characteristics is restored.
Obviously that for $H_{ext}<0$ we have the same physics, but the
excess current corresponds to the opposite case ($I>0$) and
therefore $I_c^{(+)}>I_c^{(-)}$. All these conclusions are in
agreement with our experimental observations
[Fig.~\ref{Fig-CriticalCurrents}].

Summarizing, we showed that a nonuniform field can cause a
pronounced asymmetry of the $I-V$ curves of a superconducting
bridge provided $|H_{ext}|\simeq B_0$. The difference in the
critical currents can be attributed to a removal of vortices from
the inner part of the superconducting bridge under the action of
the Lorentz force. Such a vortex--free channel, forming only for a
certain polarity of the injected current, is able to carry extra
current without dissipation and thus prevents the superconducting
bridge from switching to the normal state.

This work was supported by the Methusalem Funding of the Flemish
Government, the NES -- ESF program, the Belgian IAP, the Fund
F.W.O.--Vlaanderen, the RFBR, RAS under the Program ``Quantum
physics of condensed matter" and FTP ``Scientific and educational
personnel of innovative Russia in 2009--2013".

\end{document}